\documentclass[]{aa}
\pdfoutput=1
\usepackage{natbib,epsfig,graphicx,txfonts}
\usepackage[normalem]{ulem}
\bibpunct{(}{)}{;}{a}{}{,} 
\begin{document}
\title{Is the massive young cluster Westerlund I bound?}
\author{M. Cottaar \inst{1}
	\and M. R. Meyer \inst{1}
	\and M. Andersen \inst{2}
	\and P. Espinoza \inst{3}}
\institute{Institute for Astronomy, ETH Zurich, Wolfgang-Pauli-Strasse 27, 8093 Zurich, Switzerland
	\and European Space Agency (ESTEC), P.O. Box 299, 2200 AG Noordwijk, The Netherlands
	\and Steward Observatory, The University of Arizona, Tucson, Arizona 85721}
\date{\today}

\abstract{Westerlund I is the richest young cluster currently known in our Galaxy, making it one of the most massive clusters for which we can resolve the individual stars even in the crowded centre. This makes it an ideal target to assess whether massive clusters formed currently will remain bound or will disperse and contribute significantly to the stellar field population.}
{Here we report a measurement of the radial velocity dispersion of Westerlund I to explore whether the cluster is currently in virial equilibrium, if it is in the process of collapse or if it is expanding and dispersing into the field.}
{We obtained MIKE/Magellan high resolution optical spectra of 22 post main-sequence stars in Westerlund I for 2 or 3 epochs with a maximum baseline of about one year. Radial velocities variations between these spectra have been measured through cross correlation.}
{We calculate the velocity dispersion from the cross correlation of five yellow hypergiants and one luminous blue variable, that show little radial velocity variations between epochs and have many spectral features in common. After taking into account the effect of small number statistics and undetected binaries, we estimate the velocity dispersion for the massive stars in Westerlund I to be $2.1^{+3.3}_{-2.1}$ \, km s$^{-1}$. For several different assumptions concerning possible mass segregation and the elongation of the cluster, we find that Westerlund I is subvirial at the 90\% confidence level.}
{We can rule out that the cluster is significantly supervirial at the 97\% confidence level, indicating that Westerlund I is currently bound. This implies that Westerlund I has survived past the point where any gas expulsion has taken place and is expected to survive for billions of years.}

	
\maketitle

\section{Introduction}
Soon after its launch about twenty years ago, the Hubble Space Telescope identified many massive young star clusters in starburst and merging galaxies \citep{Hol92}. These clusters are massive enough that they appear unlikely to disperse in a Hubble time due to tidal stripping \citep{Lam05}, which means they could possibly become future globular clusters. However, violent internal or external events could cause the cluster to become unbound, in which case the cluster will dissolve into the field, significantly contributing to the stellar populations of their host galaxy. A likely candidate for such an event would be the expulsion by stellar winds or the first supernovae of the remainder of the gas cloud out of which the cluster formed. This scenario has been extensively studied both analytically \citep{Hills80, Elmegreen83,Mathieu83} and with N-body simulations \citep{Lada84,Goo06}. These studies find that the expulsion of gas lowers the potential well in which the stars reside, causing a previously virial cluster to expand. The efficacy of the gas expulsion to disrupt clusters depends on several variables, including the star formation efficiency, the timescale of the expulsion \citep{Hills80,Mathieu83} and whether the stars have achieved virial equilibrium with the potential of the surrounding gas before the gas is ejected \citep{Goo09}. Without studying the dynamical state of these clusters, we do not know which fraction of them would survive to become future globular clusters analogues and which ones will disperse into the field.

The only information on the dynamic state of these distant massive clusters, that we can currently obtain is the radial velocity dispersion \citep[e.g.][]{Bas06}. Under the assumption of virial equilibrium the one-dimensional velocity dispersion can be used to estimate the mass of the cluster, if its size is resolved. This mass estimate is called the dynamical mass. The dynamical mass can be compared to the mass derived from the luminosity of the cluster, assuming a universal IMF. Many clusters have been found to have a significantly higher velocity dispersion, than expected from the luminosity \citep{Bas06}. Several explanations have been proposed for this, including: (i) the IMF might actually be different under extreme star forming conditions \citep{Men02}, (ii) the velocity dispersion might be significantly increased due to binaries \citep{Kou08,Gie10}, or (iii) the cluster might be supervirial, for example due to a recent expulsion of gas from the cluster \citep{Goo06,Bau07}. These scenarios can be distinguished through observations of a cluster for which we can resolve the individual stars and thus measure the binary properties and the IMF.

Westerlund I is the most massive young cluster currently known in our Galaxy \citep{Por10}. Photometric observations with the NTT/SOFI in the JHKs bands \citep{Bra08,Gen11} show an elongated cluster with a mass of $4.9^{+1.8}_{-0.5} \times 10^4 M_\odot$, a half mass radius $r_{\rm hm}=1.0$ pc, and a distance of $4 \pm 0.2$ kpc. This distance is confirmed by comparing the radial velocity of the H II gas in Westerlund I with the rotation curve of the galaxy \citep{Kot07}. At this distance from the galactic centre stellar evolution and galactic tidal stripping are expected to disperse Westerlund I over $\sim 9$ Gyr, if it is currently in virial equilibrium \citep{Bau03}. Thus Westerlund I might represent a precursor of a future low-mass, metal-rich "globular cluster". However, at an age of 4.5-5 Myr \citep{Cro06,Gen11} we expect that Westerlund I might be out of virial equilibrium due to any recent expulsion of gas, which has not been converted into stars \citep{Goo06}. In this paper we describe our measurements of the radial velocity dispersion in Westerlund I to explore whether the cluster is likely to be bound or not. If Westerlund I is in virial equilibrium we expect a one-dimensional velocity dispersion of $\sim 4.6$ km s$^{-1}$ (see Sect. \ref{sec:disc}) for the mass and radius measured by \citet{Bra08} and \citet{Gen11}.

\citet{Men07,Men09} previously measured the radial velocity dispersion of Westerlund I using K-band spectra. They found a velocity dispersion of $9.2 \pm 2.5$ km s$^{-1}$ based on a sample of ten stars. This value could be severely inflated due to the presence of binaries \citep{Kou08,Gie10}. \citet{Rit09,Rit09b} have also measured the radial velocities for a large sample of members and candidate members of Westerlund I using VLT/FLAMES+GIRAFFE spectra from 8350 to 9000 \AA{} over multiple epochs to check for variability and investigate the binary properties of these massive stars.

In this paper we present multi-epoch high-resolution spectroscopy to find the velocity dispersion corrected for binaries of Westerlund I and thus constrain the dynamical state of Westerlund I.  In Sect. \ref{sec:obs} we present the observations and discuss the data reduction. We present the radial velocity variations between epochs due to binaries or atmospheric instabilities in the target stars, and calculate the velocity dispersion using the stars with low variability in Sect. \ref{sec:res}. The implications of the measured velocity dispersion on the dynamical state of Westerlund I are discussed in Sect. \ref{sec:disc}. Finally we present our conclusions in Sect. \ref{sec:conc}. 

\section{Observations \& data reduction \label{sec:obs}}
Observations were made using the MIKE spectrograph \citep{Ber03} located on the Magellan Clay telescope at the Las Campanas Observatory in June 2009, August 2009 and July 2010. MIKE has a red and a blue arm, providing two echelle spectra with a combined wavelength coverage from 3200 to 9000 \AA. A slit of 0.7 arcseconds was used, corresponding to a resolution of 53.000 ($\sim 6$ km s$^{-1}$ per resolution element). Due to the large interstellar reddening towards Westerlund I \citep[A$_{\rm V}=9.6-12.3$][]{Neg10} usable spectra are only obtained above 5000 \AA{}. We observed 22 of the brightest spectroscopically confirmed members ($I < 14.7$ mag) selected from \citet{Cla05}. These stars include two of the four red supergiants, all six yellow hypergiants, a sgB[e] star, a LBV in its cool phase and 12 OB-supergiants. All of these targets were observed for 2 or 3 epochs.

The spectra were reduced with the MIKE Redux pipeline\footnote{The MIKE Redux pipeline was written by S. M. Burles, J. X. Prochaska and R. Bernstein in IDL (http://web.mit.edu/~burles/www/).} with some minor adjustments. All images were bias subtracted using the overscan regions. Flat field images were derived from internal quartz lamp frames. The slit was in place during these measurements to ensure the variation of wavelengths with the location on the CCD is similar as in the observations. To obtain the flat field in the dark area between the orders, a diffusing glass is positioned in the optical path just downstream of the slit illuminating the area between the orders. After flat fielding we still find spurious low frequency spatial variations in the illumination of the detector above $\sim 8200$ \AA{} for all spectra taken on August 2009 and July 2010, including the internal quartz lamp spectra. These features are slightly weaker for the internal quartz lamp and are spread out by the diffusing glass, suggesting that these features are created in the optical path of the telescope. To correct these variations we calculate a smoothed flat field, which no longer contains the high frequency pixel-to-pixel variations, but only these low frequency spatial features. Although these spurious features are weaker in the flat field than in the science images, we find that they can still be roughly corrected for by dividing the flat fielded image twice by this smoothed flat field. 

The curvature of the orders along the CCD is fitted using traces along the edge of the orders obtained from internal quartz lamp images. The scattered light was fitted from the dark areas by a B-spline with 10 knots between the orders and subtracted. The long slit of MIKE (5 arcseconds) allowed us to find the sky emission lines in the part of the order not illuminated by the star and subtract these emission lines. Regularly during the nights ThAr images were taken for wavelength calibration. Two ThAr lamp images were taken at different exposure times, to maximize the range of lamp line strengths, which could be used in the wavelength calibration. Combined with the position of the orders on the CCD, these ThAr images were used to calculate a wavelength for every pixel on the image. The spectra are optimally extracted onto a common wavelength frame, allowing us to add the subsequent exposures of the same target. The noise was estimated at every pixel as the sum of the Poisson error in the observed flux and the readout noise. This calculated noise is consistent with the variations between multiple subsequent observations of the same target and corresponds to a S/N of over 100 in the red part of the spectra. The flux has been normalized using a cubic spline fit with two knots.



\begin{figure}
	\begin{center}
	\includegraphics[width=.5\textwidth]{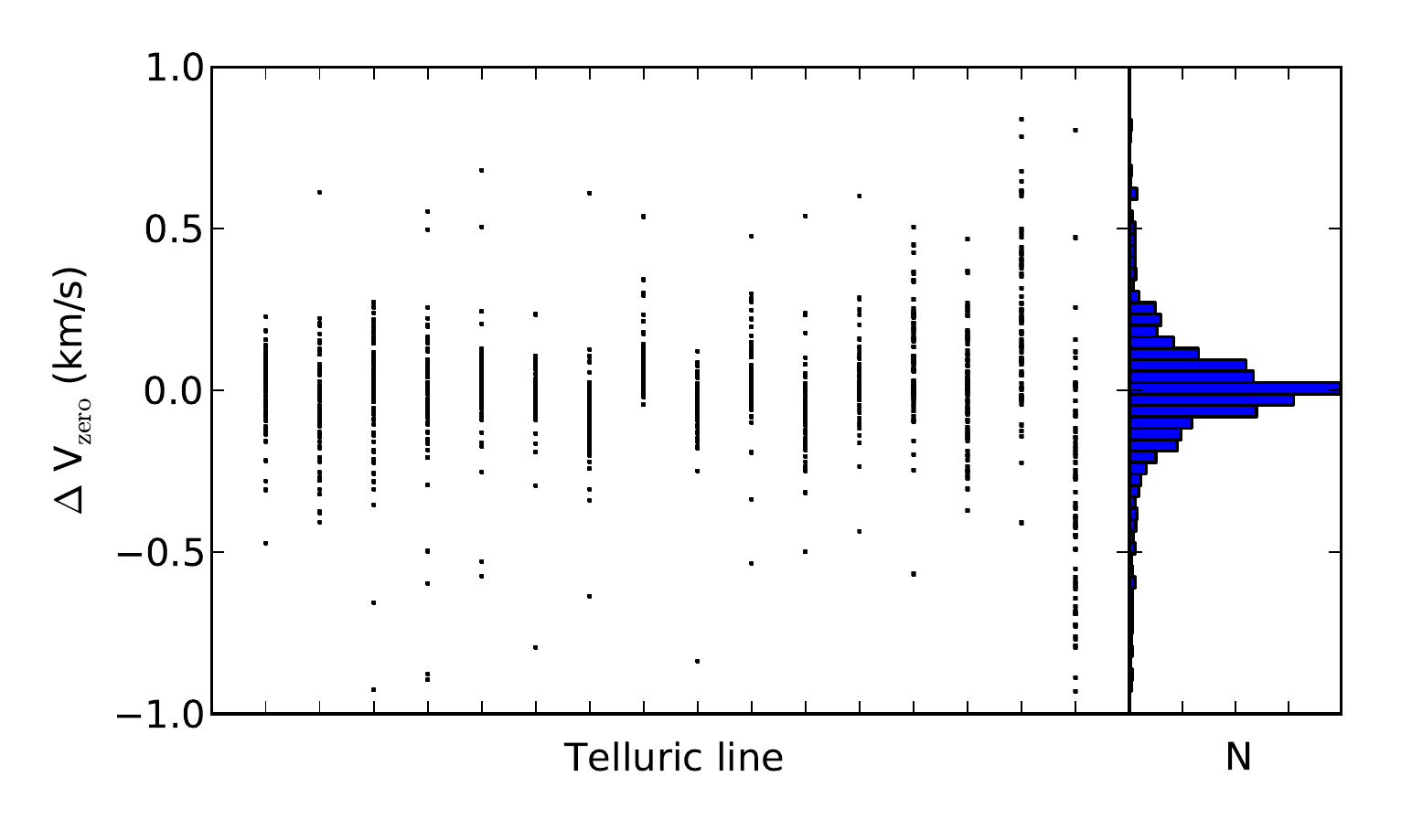}
	\caption{\label{fig:tell}The offset of the velocity zero point measured for the individual telluric lines from the averaged velocity zero point for sixteen wavelength regions containing strong telluric lines. On the right the histogram of all these offsets is plotted. The standard deviation from this distribution is about 230 m/s.}
	\end{center}
\end{figure}

To correct for possible off-center placement of the source star along the slit, we check for zero-point shifts of the wavelength solution in every observed spectrum. To this end we measure the shift of the telluric absorption lines in our stellar target spectra compared to a National Solar Observatory (NSO) telluric spectrum convolved to the resolution of our spectra. We select sixteen wavelength ranges, where there are no or only very weak contamination from stellar spectral lines or diffuse interstellar bands and which contain strong telluric lines. Over these wavelength ranges we calculate the peak of the cross correlation between the NSO spectrum and the observed spectra. The shift of the zero-point of the wavelength for every spectrum is the average of the shifts calculated for the sixteen individual wavelength ranges. Figure \ref{fig:tell} shows the difference between the individual measurements and the average offset. The distribution of these variations is fairly narrow ($\sigma =0.23$ km s$^{-1}$).  These offsets are smaller than the precision we obtain in measuring the radial velocity variations of our target stars, implying that we are not limited by the accuracy of the wavelength calibration.

\begin{table*}
	\caption{\label{tab:vel} The observed stars with the id of the star from \citet{Westerlund87} , updated by \citet{Cla05} (Col 1), its spectral type \citep{Neg10,Cla10} (Col 2), the deviation of velocities measured for a single epoch from the average velocity of that star (Col 3-5), and the maximum velocity difference between two epochs (Col 6). All velocities are in km s$^{-1}$. The relative velocities have been plotted in Figure \ref{fig:binarity}.}
	\begin{center}
	\begin{tabular}{ c  c   c  c  c  c c}
		\hline \hline
		id & sp. type & 13/14-06-2009 & 10-08-2009 & 09-07-2010 & 10-07-2010 & $\Delta {\rm v}_{\rm max}$ \\ \hline
		W56 & B0 Ia &  $-2.6 \pm 1.3$ & - & $0.1 \pm 0.8$ & $2.5 \pm 1.7$ & $5.1 \pm 2.9$\\
		W13 & B0.5 Ia & \multicolumn{5}{c}{spectroscopic binary$^{\mathrm{b}}$} \\
		W238  & B1 Iab &  $1.2 \pm 1.5$ & - & $-1.4 \pm 0.5$ & $0.2 \pm 1.1$ & $2.6 \pm 2.0$\\
		W23a & B2 Ia & $0.6 \pm 0.8$ & - & $1.9 \pm 0.7$ & $-2.5 \pm 1.3$ & $4.4 \pm 1.9$\\
		W2a & B2 Ia &  $-5.2 \pm 1.1$ &  $-1.1 \pm 0.6$ &-& $6.3 \pm 1.2$ & $11.5 \pm 2.3$\\
		W11 & B2 Ia &  $4.6 \pm 0.8$ & - & $-6.5 \pm 0.4$ & $1.9 \pm 0.8$ & $11.1 \pm 1.0$\\
		W28 & B2 Ia &  $5.7 \pm 0.8$ & $-5.7 \pm 0.8$ & - & - & $11.3 \pm 1.7$\\
		W71 & B2.5 Ia &  $-5.9 \pm 0.5$ & $5.9 \pm 0.5$ & - & - & $11.8 \pm 1.0$\\
		W70 & B3 Ia &   $-6.2 \pm 0.8$ & $6.2 \pm 0.8$ & - & - & $ 12.3 \pm 1.5$\\
		W57a  & B4 Ia &  $0.4 \pm 0.2$ & $-3.3 \pm 0.4$ & $1.1 \pm 0.5$ & $1.8 \pm 0.4$ & $5.1 \pm 0.5$\\
		W33 & B5 Ia+ &  $6.4 \pm 0.8$ &  $1.2 \pm 1.1$ & $-7.6 \pm 0.6$ & - & $14.1 \pm 0.8$\\
		W7 & B5 Ia+ &  $-4.0 \pm 0.5$ & $2.7 \pm 0.9$ & - & $1.3 \pm 0.6$ & $6.7 \pm 1.3$\\ 
		W16a & A5 Ia+ &  $4.7 \pm 0.2$ & $-1.4 \pm 0.4$ & - & $-3.3 \pm 0.5$ & $8.0 \pm 0.6$\\ 
		W243$^{\mathrm{a}}$  & LBV & $-0.5 \pm 0.3$ &  $-0.5 \pm 0.2$ & $1.0 \pm 0.3$ & - & $1.6 \pm 0.6$\\ 
		W12a$^{\mathrm{a}}$  & F1 Ia+ &  $-0.7 \pm 0.4$ & $1.2 \pm 0.2$ & - & $-0.4 \pm 0.4$ & $1.9 \pm 0.5$\\ 
		W4$^{\mathrm{a}}$  & F3 Ia+ &  $-2.4 \pm 0.2$ & $2.4 \pm 0.2$ & - & $-0.1 \pm 0.2$ & $4.8 \pm 0.4$\\ 
		W265$^{\mathrm{a}}$  & F1-5 Ia+ & $-1.2 \pm 0.3$ & $2.0 \pm 0.2$ & - & $-0.8 \pm 0.2$ & $3.2 \pm 0.5$\\ 
		W32$^{\mathrm{a}}$  & F5 Ia+ &  $-1.8 \pm 0.1$ & $1.8 \pm 0.1$ & - & - & $3.6 \pm 0.3$\\ 
		W8a$^{\mathrm{a}}$  & F8 Ia+ &  $-0.8 \pm 0.1$ & $-0.4 \pm 0.1$ & $1.1 \pm 0.1$ & - & $1.9 \pm 0.2$\\ 
		W237 & M3 Ia &  $0.7 \pm 0.6$ & $-1.0 \pm 0.5$ & $0.3 \pm 1.0$ & - & $1.7 \pm 0.5$\\
		W26  & M5-6 Ia &  $0.6 \pm 0.2$ & $-0.2 \pm 0.1$ & $-0.5 \pm 0.3$ & - & $1.1 \pm 0.5$\\
		W9 & sgB[e] &  $0.5 \pm 0.4$ & - &$-0.4 \pm 0.4$ & $-0.1 \pm 0.2$ & $0.9 \pm 0.7$\\  \hline
	\end{tabular}
	\end{center}
	\begin{list}{}{}
		\item[$^{\mathrm{a}}$]These stars have been included in our sample to calculate the radial velocity dispersion (Sect. \ref{sec:vdisp}).
		\item[$^{\mathrm{b}}$]See \citet{Ritchie10} for a discussion on the spectroscopic binary W13.
	\end{list}
\end{table*}	

\section{Results\label{sec:res}}
\subsection{Radial velocity variations \label{sec:var}}
Radial velocity variations due to binary stars or activity in the stellar atmospheres can inflate our measurement of the velocity dispersion of Westerlund I. To identify the variable stars, we search for radial velocity variations between epochs over our $\sim$1 year baseline. We calculate the velocity difference between epochs on a line-by-line basis, by cross correlating the spectra around selected spectral lines. To determine the peak of the cross correlation on a sub-pixel level we then fit a Gaussian to the 21 points closest to the peak.

For every spectral line the procedure above will give us the spectral shift between any two epochs. We write this measured spectral shift between epoch $i$ and epoch $j$ as $\Delta {\rm V}_{ij}$. If we have more than two epochs, than the best estimate of the line shift between two epochs is not only measured by the direct cross correlation between these two spectra ($\Delta {\rm V}_{ij}$), but also through the difference of the velocities measured relative to any other epoch $k$ ($\Delta {\rm V}_{ik}-\Delta {\rm V}_{kj}$). These different determinations of the spectral shifts will not be the same in general (i.e. $\Delta {\rm V}_{ij} \ne \Delta {\rm V}_{ik}-\Delta {\rm V}_{kj}$). \Citet{AllendePrieto07} showed that it is possible to combine the information of the spectral shift from the direct cross correlation and from the spectral shifts measured to any other epoch, to get a more accurate estimate of the actual spectral shift, called $\Delta V'_{ij}$ (their equation 3). These estimates will obey $\Delta V'_{ij} =\Delta V'_{ik}-\Delta V'_{kj}$, so they can be used to assign a unique velocity to every epoch, relative to some arbitrary zero-point. We will set this zero-point to the mean of the relative velocities.

\begin{figure*}
	\begin{center}
		\includegraphics[width=\textwidth]{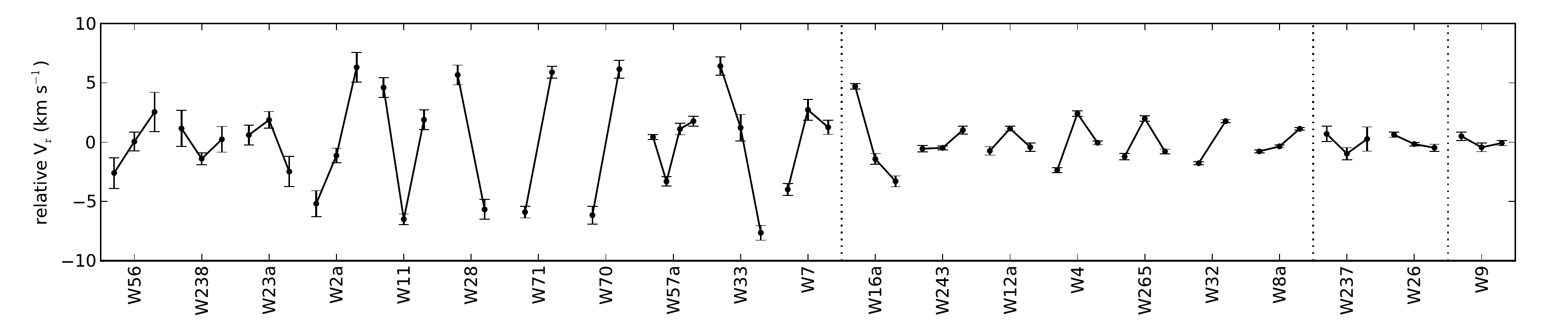}
		\caption{\label{fig:binarity} The radial velocity of an epoch relative to the mean of the velocities of all epochs that the star was observed in km s$^{-1}$ with 1-sigma uncertainties. These relative velocities have also been listed in Table \ref{tab:vel}. The vertical dotted lines separate from left to right, the B-supergiants, the YHG hypergiants (and the LBV W243), the M supergiants, and the sgB[e] star W9. Only the stars W56, W238 and W9 are consistent with having no radial velocity variations.}
	\end{center}
\end{figure*}

Using the method from \citet{AllendePrieto07} we thus get an estimate of the relative velocities between the epochs for every spectral line. Figure \ref{fig:binarity} shows this relative velocity averaged over all spectral lines for every star, except the spectroscopic binary W13 \citep{Ritchie10}. The error bars are given by the standard deviation of the mean, which will be an accurate representation of the uncertainty if it is dominated by random Gaussian errors such as expected if the noise in the spectra is the main source of uncertainty.  These velocity shifts have also been listed in Table \ref{tab:vel}, together with the largest velocity difference measured between two epochs. Only three stars (W56, W238, and W9) are consistent with having no spectral line shifts at the 2-sigma level. All four of these stars tend to have large uncertainties of a few km s$^{-1}$. For all the others the radial velocity variations between the epochs are much larger than the quoted error bars, which mean that the same spectral shift is found consistently along multiple spectral lines. Similar analysis of our radial velocity standards show a discrepancy between our estimates and the published values \citep{Udry99} of about 0.4 km s$^{-1}$. Although this is much smaller than the radial velocity changes we measure in our target stars, it is still slightly larger than our error estimate, indicating that we might underestimate our errors. Note that radial velocity changes in our target stars do not have to be caused by a binary star. Instead they may be related to stellar activity, such as large-scale super-granulation or changes in the stellar winds. That atmospheric activity can lead to radial velocity variations of many km s$^{-1}$ for these massive post-main sequence stars is well established in the literature \citep[e.g.][]{Jag98,Lobel98}.

\begin{figure}
	\begin{center}
		\includegraphics[width=.5\textwidth]{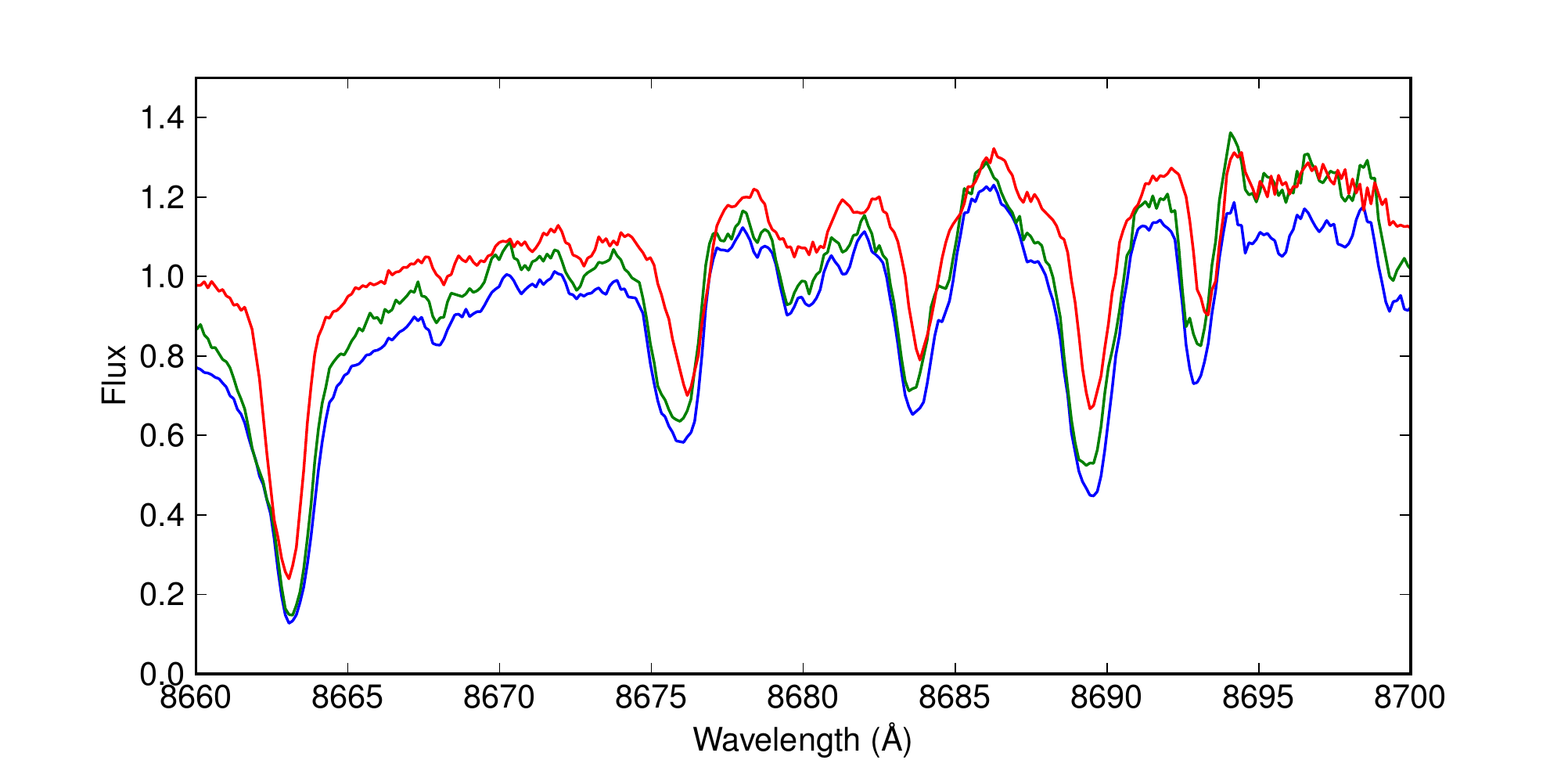}
		\caption{\label{fig:Wd237} The spectra of the three epochs of Wd1 237 (M3 Ia) with blue corresponding to June 2009, green to August 2009 and red to July 2010. The Ca II line at 8663 \AA{} is slightly blueshifted in July 2010 compared to the other epochs, while the other lines are clearly redshifted.}
	\end{center}
\end{figure}

That the cause of these spectral line shifts could be atmospheric as well as binaries is illustrated by the evolution of the spectrum of the M3 supergiant W237 shown in Figure \ref{fig:Wd237}. Over time the Ca II line at 8663 \AA{} is seen to evolve differently from most other lines. While the Ca II line is blueshifted in the last epoch (red line in Fig \ref{fig:Wd237}) compared to the earlier observations, the other lines are clearly redshifted. These inconsistent line shifts are repeated throughout the spectrum and are the cause of the relatively large error bars for the relative velocity measurements of W237 compared to the other red supergiant W26 and the yellow hypergiants in Figure \ref{fig:binarity}.

Another argument for the importance of stellar activity can be made for W11, W23a and to a lesser extent W56. These stars show a significant shift of the spectral lines over a 1 day baseline in July 2010. Although this could be explained by a short-period binary, such short-period binaries are expected to have a velocity amplitude of 100 km/s. Such a large velocity shift is not found compared to the other epoch of these stars, observed on June 2010, which could be a coincidence or caused by either a large inclination or a low mass companion. It is unlikely that these scenarios can explain the observations for \emph{all} stars, suggesting that the spectral line shifts are likely to be caused by atmospheric variability. For the other stars, distinguishing between spectral line shifts, caused by binaries or stellar activity, is difficult. However, given the high observed binary fraction among massive stars, we do expect a significant portion of the radial velocity variations to be caused by binary companions \citep{Sana11}.

The stars in Table \ref{tab:vel} and on the x-axis of Figure \ref{fig:binarity} have been roughly sorted by spectral type. The LBV is currently in a cool phase and has a lot of lines in common with the A/F hypergiants. Figure \ref{fig:binarity} shows two trends with spectral type. The first trend is a decrease in the error bars to later spectral types, which is caused by the far larger sample of spectral lines, most of which are less broad, which are available for the later type stars. The main exception on this trend is W237, which has been discussed above (see Fig \ref{fig:Wd237}). The other trend is that the yellow hypergiants and the red supergiants tend to show lower radial velocity variations than the B-supergiants. This variation stays much larger than the error bars, which means that this trend is consistently found over many different spectral lines. This could be explained by the fact that we are sensitive to relatively close binaries with periods of at most a few years. \citet{Eld08b} showed that yellow hypergiants and red supergiants in close binaries evolve more quickly into a Wolf-Rayett star, because they lose their hydrogen atmosphere more effectively in binary-induced mass loss. So yellow hypergiants and red supergiants in close binaries spend less time in these extended phases before becoming a Wolf-Rayett star than if these stars were in wide binaries or in isolation \citep{Cla11}. This creates a selection effect, where observed yellow hypergiants or red supergiants tend to have a low fraction of close binaries.

\subsection{Velocity dispersion \label{sec:vdisp}}
We calculate the velocity dispersion of the cluster using only the yellow hypergiants and the LBV. This sample was chosen because these stars show a small amount of  radial velocity variations compared to most other stars and they have a large number of spectral lines in common with each other. The large number of lines in common means that we can measure the velocity dispersion on the basis of the cross correlation without having to deal with the systematics, which complicate the measurement of the absolute radial velocity. When trying to include more low variability stars in this sample, we find that the extra noise caused by having to place the radial velocities on an absolute scale offsets the advantage that we gain by having a larger sample size. The choice to base our measurement of the velocity dispersion on the cross correlation implies that we a direct comparison with previous radial velocity measurements of the stars in Westerlund I is not possible \citep{Men07,Men09,Rit09,Rit09b}.

\begin{figure}
	\begin{center}
		\includegraphics[width=.5\textwidth]{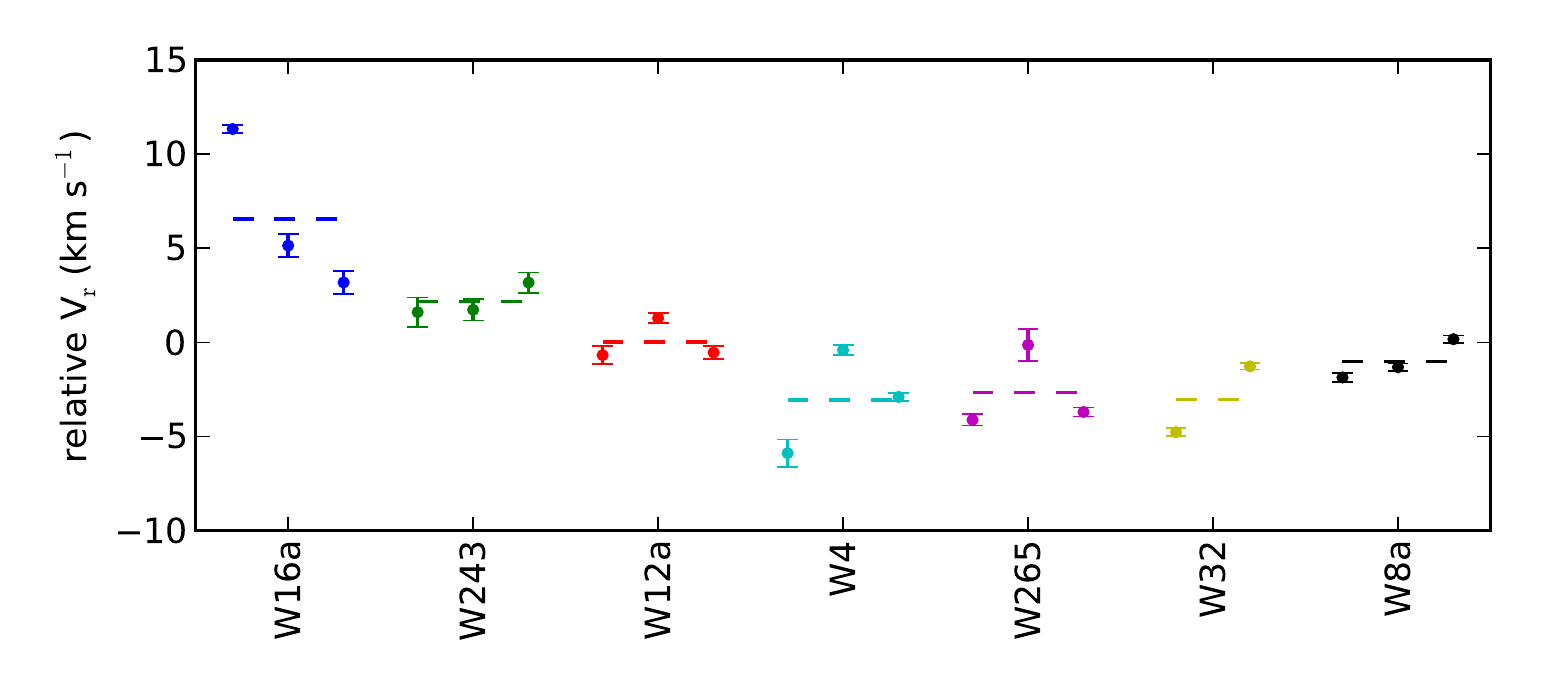}
		\caption{\label{fig:cc_vdisp} The velocities of all epochs of the YHG and the LBV relative to each other. The horizontal dashed lines show the mean velocity for every star. The zero-point of the y-axis has been arbitrarily set to the mean of all the velocities. }
	\end{center}
\end{figure}

Using the same method as in our search for radial velocity variations, we use cross correlations to calculate the relative velocities between the different stars in our sample. The relative velocities for all the epochs of the yellow hypergiants is shown in Figure \ref{fig:cc_vdisp} with the mean relative velocity for every star marked as a dashed line. We see that the velocity variations between epochs of the same star is on the same order as the velocity differences between stars. This implies a very low intrinsic velocity dispersion for Westerlund I. Based on the outlying average velocity and the large velocity change between epochs we identify W16a as a binary star (or as having a very active atmosphere) and exclude it from our calculation of the velocity dispersion. Below we calculate the velocity dispersion based on the other six stars.

\begin{figure}
	\begin{center}
		\includegraphics[width=.5\textwidth]{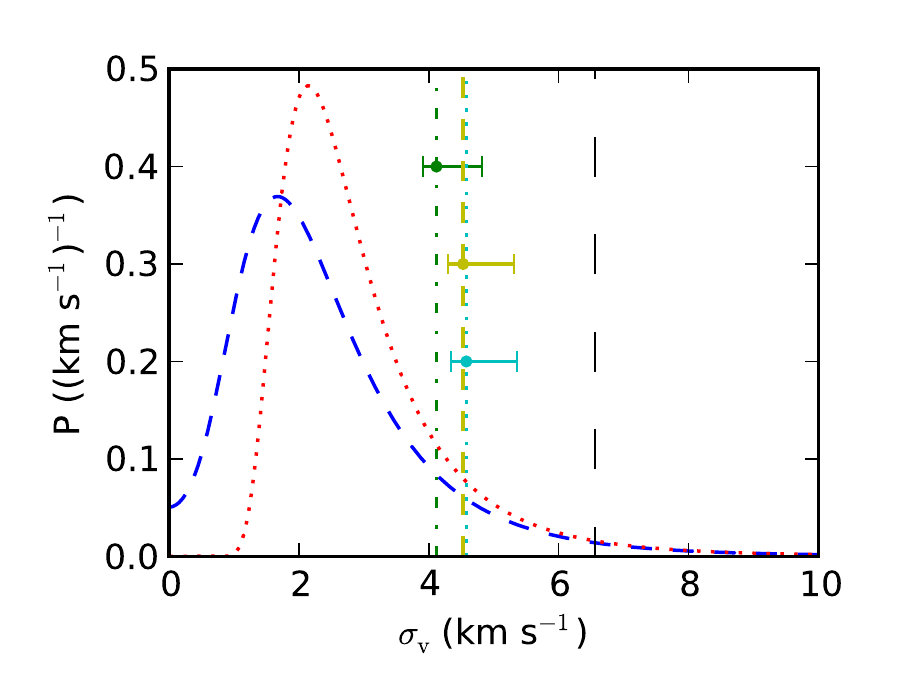}
		\caption{\label{fig:p_vdisp} Our estimate of the probability distribution of the velocity dispersion for the massive stars in Westerlund I. The blue dashed line shows the case where the measurement error is is set by the variation of the radial velocity between epochs. The red dotted line shows the limit, where we assume no measurement errors. The vertical lines mark the velocity dispersion expected in virial equilibrium under several assumptions explained in Sect. \ref{sec:disc}. From the highest to lowest estimates we have the velocity dispersion calculated for a spherical, non-mass segregated cluster for the mass estimates from \citet{Neg10} (black, long dashed) and \citet{Gen11} (cyan, dotted). For the mass estimate of \citet{Gen11} we also show the expected velocity dispersion taking into account the velocity anisotropy, suggested by the elongation of Westerlund I (yellow, dashed), and the case where we take into account both the velocity anisotropy and the possible mass segregation (green, dash-dot). The error bars represent the quoted uncertainty in the photometric mass of \citet{Gen11}.}
	\end{center}
\end{figure}

Due to the small number of stars in our sample we will have a large statistical uncertainty in the measured velocity dispersion. To calculate this statistical uncertainty we calculate the likelihood that the observed distribution would have been drawn from a cluster with a Gaussian radial velocity distribution with standard deviation $\sigma_{\rm v}$ and mean $\mu$:
\begin {equation}
	P(\sigma_{\rm v},\mu)=\displaystyle\prod_i \frac{1}{\sqrt{2 \pi (\sigma_{\rm v}^2+\sigma_i^2)}} \exp{\left(-\frac{(v_i-\mu)^2}{2(\sigma_{\rm v}^2+\sigma_i^2)}\right)},
\end{equation} 
where $v_i$ is the observed mean velocity of the star and $\sigma_i$ is the uncertainty on the measured velocity. The likelihood that the observations were drawn from a distribution with a given velocity dispersion is calculated by integrating over $\mu$. With only 2-3 epochs over a 1 year baseline and most, if not all, stars being variable the uncertainties on the measured radial velocities of the stars ($\sigma_i$) are not well determined.  Because of this we will consider two cases. In the first case we set $\sigma_i^2$ to the variance of the velocity spread between epochs. This leads to the probability distribution shown as the blue dashed line in Figure \ref{fig:p_vdisp} and corresponds to an intrinsic velocity dispersion of $\sigma=1.7^{+3.1}_{-1.7}$ km s$^{-1}$, where the errors give the 95\% confidence limits. Alternatively we set $\sigma_i$ to zero, leading to a velocity dispersion of  $\sigma=2.1^{+3.4}_{0.9}$. This case is shown as the red dotted line in Figure \ref{fig:p_vdisp}. Note that the peak in the second case overlaps exactly with the standard deviation of the measured stellar velocities (2.15 km s$^{-1}$).

The second case, with $\sigma_i=0$, can be seen as an "upper-limit estimate"  of the probability distribution of true velocity dispersion of Westerlund I. This velocity dispersion would be accurate if the errors were much smaller than 1 km s$^{-1}$. Given the large radial velocity variations measured between epochs, as well as the possibility of undetected binaries with a period of $>5$ years and a velocity amplitude of tens of km s$^{-1}$, it is a very unlikely scenario that the uncertainties in our estimate of the systematic radial velocity are below 1 km s$^{-1}$. However, including more accurate measurement errors would only decrease the measured velocity dispersion, as illustrated by the case where we set the measurement errors to the variance of the velocity variations between epochs (blue dashed line in Figure \ref{fig:p_vdisp}). Increasing $\sigma_i$ does not change the probability at the higher velocity dispersions significantly, i.e. the upper bound of the 95\% confidence limit remains around 5 km s$^{-1}$. Because the uncertainties are not well defined, we will conservatively focus on the case with $\sigma_i=0$.

\section{Discussion \label{sec:disc}}
To explore the virial state of the cluster, we compare our measurement of the velocity dispersion with the velocity dispersion expected in virial equilibrium for the photometric mass of Westerlund I. We calculate a first estimate of the velocity dispersion expected in virial equilibrium assuming Westerlund I is not mass segregated and that the velocities are isotropic. In virial equilibrium the photometric mass is equal to the dynamical mass, which under these assumptions is given by:
\begin{equation}
	M_{\rm dyn}=\eta \frac{\sigma_{\rm rad}^2 r_{\rm hm}}{G}, \label{eq:mdyn}
\end{equation}
where $\eta \approx 10$ for the density profile found by \citet{Bra08}, $\sigma_{\rm rad}$ is the radial velocity dispersion, and $r_{\rm hm} \approx 1.0 pc$ \citep{Bra08} is the half mass radius. Based on an extrapolation from the number of massive stars in Westerlund I, \citet{Neg10} estimated a stellar mass of about $10^5 \, {\rm M}_\odot$. In virial equilibrium this corresponds to a velocity dispersion of $\sim 6.5$ km s$^{-1}$. This is far higher than our measured velocity dispersion, implying that Westerlund I is subvirial at the 98 \% confidence level. An alternative photometric mass estimate has been given by \citet{Gen11}. Based on an extrapolation of the measured IMF for the intermediate mass stars between 3.5 and 27 M$_\odot$, they found a lower stellar mass of  $4.9^{+1.8}_{-0.5} \times 10^4 \, {\rm M}_\odot$. In virial equilibrium we would then expect a velocity dispersion of $4.6^{+0.8}_{-0.3}$ km s$^{-1}$. For this velocity dispersion our measurements still favor a subvirial cluster at the 91 \% confidence level.

In the conversions from the photometric mass to the velocity dispersion calculated above we have assumed that the cluster has an isotropic velocity distribution and that it is not mass segregated. Both of these assumptions are unlikely to be valid for Westerlund I. An isotropic velocity distribution would generally lead to a spherical cluster. However, Westerlund I is observed to be elongated, with surfaces of constant density lying on ellipses rather than circles \citep{Gen11}. The exact implications of this elongation on the expected velocity dispersion depend on its origins. Under the assumption that this elongation is supported by a velocity anisotropy and thus is not a transient on the dynamical timescale, we calculate in appendix \ref{app:ell} the expected velocity dispersion. We find that the dynamical mass is given by eq \ref{eq:ell_mdyn}, reproduced here:
\begin{equation}
	M_{\rm dyn}=\eta_{\rm ellips} \eta\frac{\sigma_{\rm rad}^2 r^\prime_{\rm hm}}{G}. \label{eq:mdyn_ell}
\end{equation}
where $r^\prime_{\rm hm}$ is the half mass radius, if the density profile in all dimensions was the same as along the semi-major axis. $r^\prime_{\rm hm}$ can be estimated from the half-mass radius of 1.0 pc \citep{Bra08} and the ellipticity of Westerlund I of 0.75 \citep{Gen11} to be roughly 1.2 pc. The correction factor for the ellipticity of the cluster $\eta_{\rm ellips}$ is approximately 0.85, if the axis along the line of sight is not much smaller than the observed axes. The other parameters in the equation are the same as in eq \ref{eq:mdyn}. For the photometric mass estimate of $4.9^{+1.8}_{-0.5} \times 10^4 {\rm M}_\odot$ \citep{Gen11} taking into account the elongation leads to a only slightly lower expected velocity dispersion of $4.5^{+0.8}_{-0.2}$ km s$^{-1}$.

Due to the tendency of massive stars to reside in the center of massive clusters (i.e. mass segregation) the velocity dispersion of the massive stars which we observe might not representative of the cluster as a whole \citep{Fle06}. To calculate the effect of mass segregation on the difference between the measured and expected velocity dispersion, we separate the cluster in two populations. The first population represents the massive post-main sequence stars from which our sample was drawn, with a half-mass radius of $0.44$ pc and a mass of $6 \times 10^3 M_\odot$ \citep{Cla05,Neg10}. The second population contains the remaining low and intermediate mass stars in the cluster, with a half-mass radius of $1.0$ pc and a much higher total mass of $4.3 \times 10^4 M_\odot$ \citep{Bra08}. For a cluster whose stellar volume density is well-represented by a Plummer sphere we find that in virial equilibrium the velocity dispersion of the total population is about 10\% larger than for the subset of massive star using a two-fluid approximation \citep[eq 7.158 in][]{Bin08}. This would lead us to expect to measure a velocity dispersion of roughly $4.1^{+0.7}_{-0.2}$ km s$^{-1}$ for a mass segregated, elongated cluster.

Besides showing the probability distribution of the velocity dispersion of the massive stars in Westerlund I implied by our observations, Figure \ref{fig:p_vdisp} also illustrates the velocity dispersions expected in virial equilibrium using the various assumptions discussed above. Even for the lowest estimate our observations still strongly favor that Westerlund I is subvirial or virial. For the probability distribution including a rough estimate of our measurement errors we find a 91\% probability of the observed velocity dispersion to lie below the lowest estimate of 4.1 km s$^{-1}$. For the more conservative probability distribution, which assumes no measurement errors, we still find a 87\% probability.


For the cluster to be unbound, the kinetic energy has to be higher than the binding energy, which would correspond to a velocity dispersion $\sqrt{2}$ higher than expected in virial equilibrium. This can be excluded with at least 97 \% confidence from our measurements, with the exact percentage depending on the assumptions made. This implies that Westerlund I is bound and has survived any recent gas expulsion. To survive the gas expulsion Westerlund I should either have had a high star-formation efficiency, which would cause the cluster to remain close to virial equilibrium or the stars in the cluster should have been dynamically cold at the moment the gas was expelled \citep{Goo09}. If the case of Westerlund I is typical for a massive cluster, this would mean that the excessive velocity dispersion measured for extragalactic clusters does not automatically imply a supervirial cluster, but has another source, for example binaries \citep{Gie10}. In any case, Westerlund I itself will not disperse into the field quickly and will likely persist for billions of years.

\section{Conclusions\label{sec:conc}}
We present multi-epoch high resolution spectra of 22 massive post main-sequence stars in the young, massive, Galactic cluster Westerlund I. For 21 of these 22 stars the velocity variations have been measured between multiple epochs by cross correlation (see Table \ref{tab:vel} or Fig \ref{fig:binarity}). Many of these stars seem to show radial velocity variations at the one to ten km s$^{-1}$ level. Some of these variations might be explained by an underestimate of our uncertainties. Alternatively these variations could be caused either by binaries or by instabilities/waves in the complex atmospheres of these super- and hypergiants.

Using the five least variable yellow hypergiants and the LBV, we find a velocity dispersion for the massive stars in Westerlund I of $\sigma=2.1^{+3.4}_{-0.9}$ km s$^{-1}$ with a confidence limit of 95\%, assuming no significant errors in measured radial velocity differences between the stars. In reality even these least variable stars show radial velocity variations between epochs of a few km s$^{-1}$, implying that the quoted velocity dispersion is only an upper limit of the true velocity dispersion. 

In Figure \ref{fig:p_vdisp} the probability distribution of the velocity dispersion from our measurements is compared to several estimates of the velocity dispersion which we expect in virial equilibrium. These estimates have been calculated using different estimates of the total mass of Westerlund I, as well as different rough approximations of the effects of possible mass segregation and the effects of a velocity anisotropy, which is suggested by the observed spatial elongation of Westerlund I. For all of these cases the observations favor a subvirial cluster. For the lower photometric mass estimate of \citet{Gen11} the observations are consistent with the cluster in virial equilibrium. In all cases we exclude that Westerlund I is significantly supervirial, implying that the cluster is bound, has survived any recent gas expulsion. Barring any violent interactions with other clusters or molecular clouds, Westerlund I, which might be the most massive young cluster in our Galaxy, is expected to survive for billions of years.

\begin{acknowledgements}
NSO/Kitt Peak FTS data used here were produced by NSF/NOAO. We would like to thank Justin Read, Kevin Covey, Richard Parker, Phil Massey and Hans Martin Schmidt for helpful discussions. We are also very grateful to an anonymous referee, whose comments and suggestions led to major improvements in the paper.
\end{acknowledgements}

	\bibliographystyle{aa}
	\bibliography{west,global}
\appendix
\section{Velocity dispersion in an ellipsoidal cluster\label{app:ell}}
The density distribution of both the massive and intermediate mass stars in Westerlund I has been shown to be elongated \citep{Neg10,Gen11}. Here we will calculate the effect of this elongation on the dynamical mass calculation of Westerlund I, using an ellipsoidal density profile. Such a profile is suggested by \citet{Gen11}, who fitted a density profile to their stellar density distribution of Westerlund I, where the lines of constant densities lie along ellipses instead of circles. To limit the number of free parameters as well as greatly simplify the equations, we will assume that one of the axis of the ellipsoid lies along the line of sight. The other two axes are then the axes found by \citet{Gen11} in the plane of the sky. We define the length of the semi-major axis found by \citet{Gen11} as $a_1$, the length of the semi-minor axis as $a_2$ and than have $a_3$ as the length along the line of sight, which is a unconstrained by observations.

For an ellipsoidal cluster in virial equilibrium the elongation implies a velocity anisotropy, which is given by the tensor form of the virial theorem
\begin{equation}
	2 K_{jj}+W_{jj}=0, \label{eq:ell_vir}
\end{equation}
where $2 K_{jj}=M_{\rm dyn} \sigma_j^2$ is the tensor form of the kinetic energy along the axis $j$ and $W_{jj}$ is the potential energy tensor. For an ellipsoidal system this potential can be split up in a product of two factors \citep[see eq 2.144 in][]{Bin08}: (i) which only depends on the axial ratios of the cluster; and (ii) which is independent of the cluster's ellipticity or along which axis the potential energy tensor is calculated and only depends on the density dropoff along a single axis, which we take to be the observed semi-major axis. In functional form this corresponds to
\begin{equation}
	W_{jj}=-f_{j}(a_1,a_2,a_3)*g(\rho_{\rm along\:semi-major\:axis}), \label{eq:ell_pot}
\end{equation}
with
\begin{equation}
	f_{j}(a_1,a_2,a_3)=\frac{3}{2}\frac{a_2 a_3}{a_1^2}\left(\frac{a_j}{a_1}\right)^2 A_j, \label{eq:ell_ratio}
\end{equation}
where the $A_j$ are a complex function of the axial ratios given by Table 2.1 in \citet{Bin08}. Note that the equations in this table assume $a_1>a_2>a_3$, which will make it necessary to rearrange the indices, if the axis along the line of sight is not the smallest. 

\begin{figure}
	\begin{center}
		\includegraphics[width=.5\textwidth]{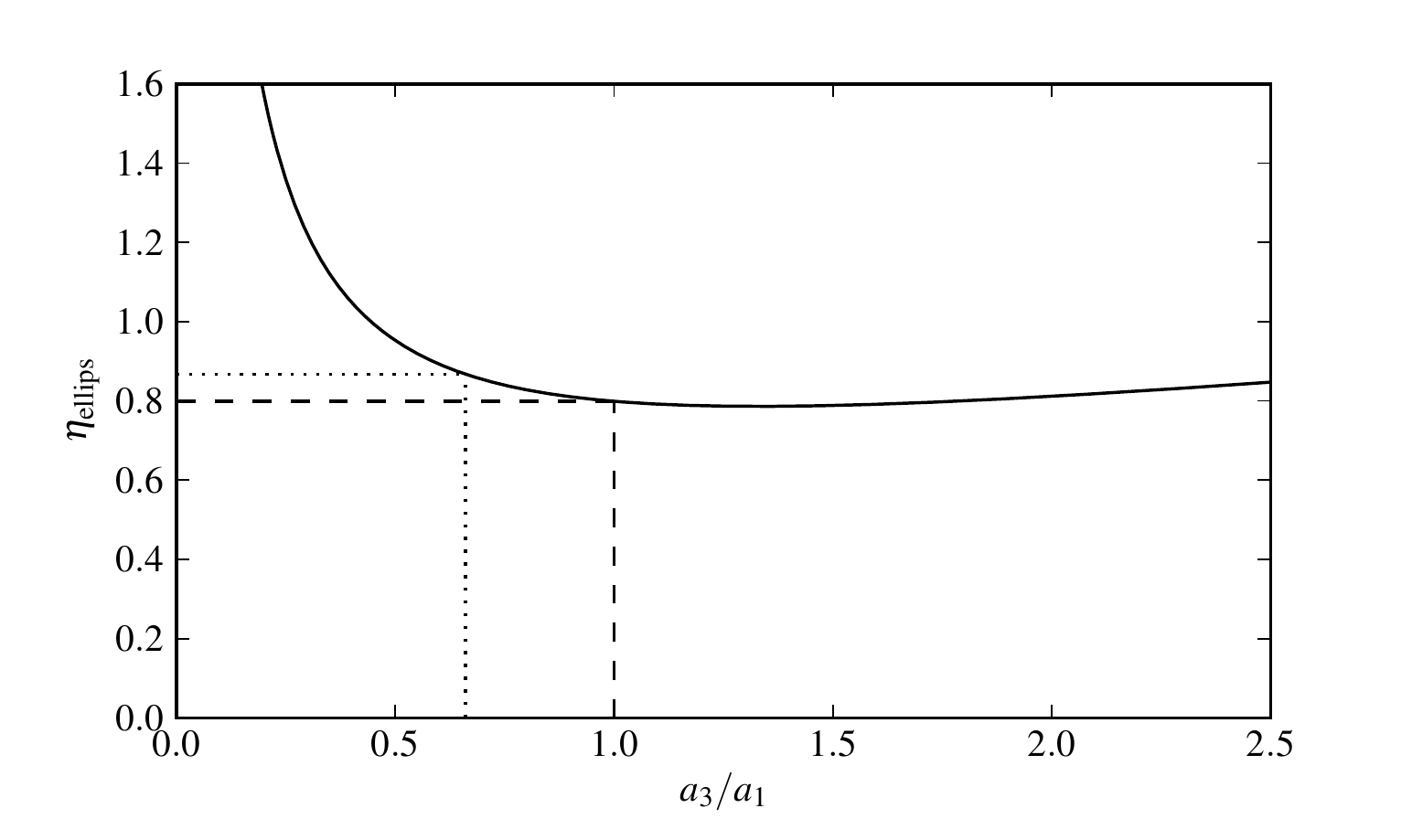}
		\caption{\label{fig:ell_corr} The correction factor $\eta_{\rm ellips}$ needed to take into account the spheroidal shape of the cluster plotted against the ratio of the length of the axis along the line of sight to the length of the observed semi-major axis for an observed ellipticity of 0.75 ($a_2/a_1 \approx 0.66$). The cases where the density dropoff along the line of sight is the same as the semi-major axis (dashed) and the same as the minor axis (dotted) are shown.}
	\end{center}
\end{figure}

Because $g$ does not depend on the $a_2$ and $a_3$ we can calculate this term for the spherical case, where we set $a_2$ and $a_3$ to $a_1$. For a spherical cluster we have $A_j=2/3$, so $f_{j}(a_1,a_2,a_3)=1$ and $W_{jj}=\frac{1}{3}W$, where $W$ is the total potential energy. Using the equation for the dynamical mass of a spherical cluster \citep[e.g.][]{Por10} we get
\begin{equation}
	W_{jj,\rm{circ}}=-g=-\frac{1}\eta \frac{G M_{\rm dyn}^{\prime 2}}{r^\prime_{\rm hm}}, \label{eq:W_circ}
\end{equation}
where $\eta \approx 10$ \citep{Por10} for the density profiles fitted by \citet{Bra08} and \citet{Gen11}, $M_{\rm dyn}^{\prime}$ is the mass of this circular cluster and $r^\prime_{\rm hm}$ is the half mass radius of this cluster. The dynamical mass of this artificial spherical cluster is related to the mass of the spheroidal cluster with
\begin{equation}
	\frac{M_{\rm dyn}^\prime}{M_{\rm dyn}}=\frac{a_1^3}{a_1 a_2 a_3}. \label{eq:mass_ratio}
\end{equation}

Entering equations \ref{eq:ell_ratio}-\ref{eq:mass_ratio} in the equation for the ellipsoidal potential energy tensor gives
\begin{equation}
	W_{jj}=-\frac{3}{2}\frac{a_j^2}{a_2 a_3} A_j \frac{1}{\eta} \frac{G M_{\rm dyn}^2}{r^\prime_{\rm hm}}.
\end{equation}
Combined with the virial equation \ref{eq:ell_vir} this can be converted for the radial velocity dispersion ($j=3$) as:
\begin{equation}
	M_{\rm dyn}=\eta_{\rm ellips} \eta \frac{\sigma_{\rm rad}^2 r^\prime_{\rm hm}}{G}, \label{eq:ell_mdyn}
\end{equation}
where the correction for the ellipsoidal structure has been completely integrated in $\eta_{\rm ellips}$, which is defined as
\begin{equation}
	\eta_{\rm ellips} \equiv \frac{2}{3} \frac{a_2}{a_3} \frac{1}{A_3}. \label{eq:ell_corr}
\end{equation}
This correction factor only depends on the axial ratios. For an ellipticity in the line of sight of 0.75 ($a_2/a_1 \approx 0.66$) \citep{Gen11}, we plot the dependence of this correction factor on $a_3/a_1$ in Fig. \ref{fig:ell_corr}. This figure shows that there is only a small dependence of the correction factor on the actual length of the axis along the line of sight for $a_3/a_1 > 0.5$. In this range an increase in the length of the axis, causes the total potential and thus the total kinetic energy to drop. However, this drop in kinetic energy mostly corresponds to a lowering of the velocity dispersion in the plane of the sky, with the radial velocity dispersion staying roughly constant. This interplay between the change in the total kinetic energy and the division of this kinetic energy over the three axes produces the curve shown in Fig. \ref{fig:ell_corr}.

\end{document}